\documentclass[twocolumn]{aastex61}

\begin{document}

\shorttitle{Sense of rotation of A}
\shortauthors{Pol {\em et al.}}

\title{A direct measurement of sense of rotation of PSR J0737--3039A}

\author{Nihan Pol}
\author{Maura McLaughlin}
\affil{Department of Physics and Astronomy, West Virginia University, Morgantown, West Virginia 26506, USA}
\affil{Center for Gravitational Waves and Cosmology, West Virginia University, Chestnut Ridge Research Building, Morgantown, West Virginia 26505}
\author{Michael Kramer}
\affil{Max-Planck Institut f\"ur Radioastronomie, Auf dem H\"ugel 69, D-53121 Bonn, Germany}
\author{Ingrid Stairs}
\affil{Department of Physics and Astronomy, University of British Columbia,
       6224 Agricultural Road, Vancouver, BC V6T 1Z1, Canada}
\author{Benetge B. P. Perera}
\affil{Jodrell Bank Centre for Astrophysics, School of Physics
and Astronomy, The University of Manchester, Manchester M13
9PL, UK}
\author{Andrea Possenti}
\affil{INAF--Osservatorio Astronomico di Cagliari, Loc. Poggio dei Pini, I-09012 Capoterra (CA), Italy}

\email{nspol@mix.wvu.edu}

\begin{abstract}
	
	We apply the algorithm published by \citet{Liang2014} to describe the Double Pulsar system J0737--3039 and extract the sense of rotation of first born recycled pulsar PSR J0737--3039A. We find that this pulsar is rotating prograde in its orbit. This is the first direct measurement of the sense of rotation of a pulsar with respect to its orbit and a direct confirmation of the rotating lighthouse model for pulsars. This result confirms that the spin angular momentum vector is closely aligned with the orbital angular momentum, suggesting that kick of the supernova producing the second born pulsar J0737--3039B was small.
	
\end{abstract}

\keywords{pulsars: general ---
          pulsars: individual (PSR J0737--3039A)}
          
\nopagebreak

\section{Introduction}
	
	The Double Pulsar, PSR J0737--3039 \citep{Burgay_discovery_2005, Lyne_Bdiscovery_2004} is the first and only neutron star binary system that has had two detectable radio pulsars. The recycled PSR J0737--3039A (hereafter `A') has a period of 22.7 ms and the younger, non-recycled PSR J0737--3039B (hereafter `B') has a spin period of 2.8~s. The 2.45-hour orbit makes this system the most relativistic binary known, providing a unique laboratory to conduct the most stringent tests of Einstein's theory of general relativity in the strong-field regime \citep{Kramer_GRtest_2006}.
	
	In addition to strong-field tests of gravity, the Double Pulsar also offers a unique laboratory to test plasma physics and magnetospheric emission from pulsars \citep{Breton_eclipse_I, emission_height, Lyutikov_eclipses}. We originally were able to detect bright single pulses from B in two regions of its orbit, $190^{\circ} \sim 230^{\circ}$, referred to as bright phase I (BP I) and $260^{\circ} \sim 300^{\circ}$, referred to as bright phase II (BP II) \citep{Lyne_Bdiscovery_2004, Perera_evolution_2010}.  \citet{Maura_mod_2004} discovered drifting features in the sub-pulse structure from pulsar B. They showed that this phenomenon was due to the direct influence of the magnetic-dipole radiation from A on B. These drifting features (henceforth referred to as the `modulation signal') are only visible in BP I when the electromagnetic radiation from A meets the beam of B from the side \citep{david_lyutikov_modelling}. These modulation features were observed to have a frequency of $\approx 44$ Hz which suggests that this emission is not from the beamed emission of A, which has a frequency of $\approx 88$ Hz due to the visibility of emission from both the magnetic poles of A.
	
	\citet{Freire_model_2009} proposed a technique to measure, among other things, the sense of rotation of A with respect to its orbit using the time of arrival of pulsed radio emission from A and the modulation feature from B. A complementary technique was proposed by \citet{Liang2014} (henceforth LLW2014) to uniquely determine the sense of rotation of A using an approach based on the frequency of the modulation signal. LLW2014 argued that we should be able to observe an effect similar to the difference between solar and sidereal periods observed in the Solar System in the Double Pulsar. Thus, if pulsar A is rotating prograde with respect to its orbit, the modulation signal should have a period slightly greater than that if it were not rotating, and it would have a slightly smaller period for the case of retrograde motion. LLW2014 provide an algorithm to apply this concept to the observations of the Double Pulsar, and we refer the reader to that paper for more details on the calculations and details of the algorithm.
	
	In this paper, we implement this algorithm on the Double Pulsar data and determine the sense of rotation of pulsar A. In Sec.~\ref{Procedure}, we briefly describe the data used and the implementation of the algorithm from LLW2014. We present our results in Sec.~\ref{Results} and we discuss the implications of these results in Sec.~\ref{Discussion}.

\section{Procedure} \label{Procedure}
	
	\subsection{Observations and Data Preparation} \label{obs}
		
		We have carried out regular observations of the Double Pulsar since December 23, 2004 (MJD 52997) with the Green Bank Telescope (GBT). The radio emission of B has shown a significant reduction in flux density ($0.177 \ \textrm{mJy} \ \textrm{yr}^{-1}$) and evolution from a single peaked profile to a double peaked profile due to relativistic spin precession with B's radio emission disappearing in March 2008 \citep{Perera_evolution_2010}. As a result, we choose the data where B's emission is brightest, which also corresponds to the modulation signal being the brightest, for this analysis, i.e. the data collected on MJD 52997.
		
\begin{figure}
			\includegraphics[width = \linewidth, clip]{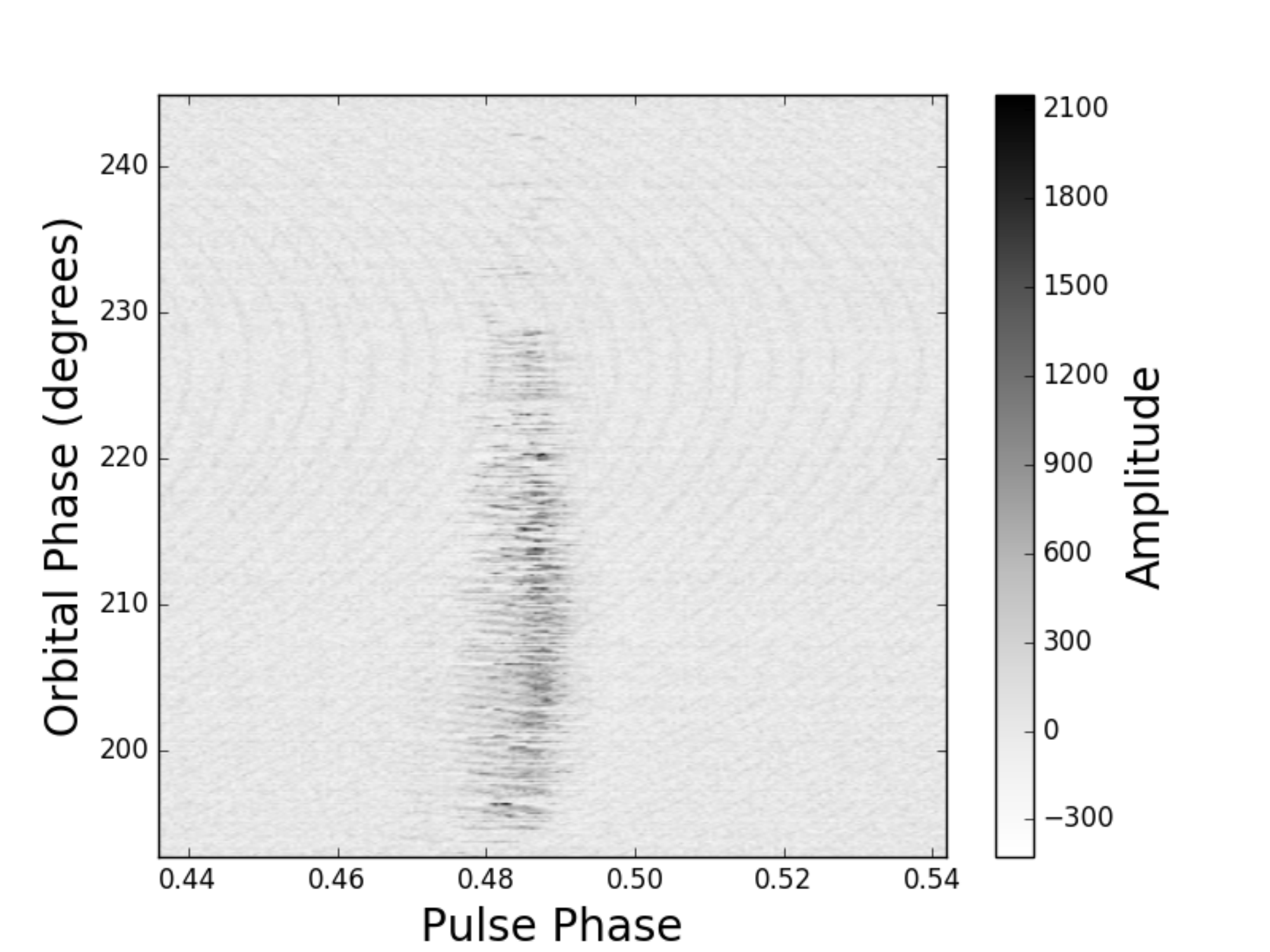}
			\caption{Single pulses of B for MJD 52997 for orbital phase $190^{\circ} - 240^{\circ}$ in the first BP I. The drifting features are most prominent in the orbital phase range of $195^{\circ} \sim 210^{\circ}$. A's pulses are also visible in the background, and are most visible at $\sim 225^{\circ}$. Note that only a fraction of the pulse phase is shown for clarity and the units on the amplitude are arbitrary. This figure is adapted from \citet[][]{Maura_mod_2004}.}
			\label{original_signal}
		\end{figure}

		These data on MJD 52997 were taken at a center frequency of $800$ MHz with the GBT spectrometer SPIGOT card \citep[][]{Spigot_ref}. This observation had a sampling time of $40.96 \ \mu$s and the observation length was 5 hours, covering more than two complete orbits. We barycenter these data using the barycenter program from the SIGPROC\footnote{This software can be found at: \url{http://sigproc.sourceforge.net/}} software package. We decimate the data from its native resolution of $24.41$ kHz to $2048 \times f_{B,0} \approx 738.43$ Hz where $f_{B,0}$ is the rest-frame frequency of B. This is equivalent to splitting up a single rotational period of B into 2048 bins. We do this to increase the signal-to-noise ratio (S/N) of the modulation signal. Since B's drifting pulses are observed only in BP I, we focus only on these orbital phases. Since this data set covers more than two orbits, we obtain two such BP I time series. The first of these BP I time series is shown in Fig.~\ref{original_signal}.
		
		Since the drifting pulses are seen only at the beginning of this phase range, we analyze orbital phase range (defined as the longitude from ascending node, i.e. the sum of longitude of periastron and true anomaly) $195^{\circ}$ to $210^{\circ}$. This final data set is approximately 344 seconds long. With the data set prepared as described above, and noting that the time series length, $T = N \Delta t$ where $N$ is total number of samples and $\Delta t$ is the sampling time, our Fourier spectra have a frequency resolution:
		\begin{equation}
			\displaystyle f_s = \frac{1}{T} = \frac{1}{N \Delta t} = 2.93 \textrm{ mHz}
		\end{equation}
		
	\subsection{Transformation}
		
		We apply the algorithm from LLW2014 (see Sec.~3.4 therein) to the time series. We found a typographic error in the algorithm from LLW2014. We describe this error and how we fixed it in Appendix~\ref{errata}. All programming was done in Python. We wrote a function that would return values for $\omega_B$ (longitude of periastron of B's orbit) and $\theta$ (true anomaly for B) as a function of time \citep[see Chapter 8 in][]{handbook}. All orbital parameters such as eccentricity $e$, orbital inclination angle $i$ and semi-major axis $a_B$ were obtained from the timing solution of B \citep{Kramer_GRtest_2006}. With all these parameters in place, the implementation of the algorithm was straight-forward. For completeness, we briefly list the transformation described in LLW2014, and refer the reader to that paper for more details.
		
		The basic idea of the transformations is to remove the Doppler smearing produced by eccentric orbits in the Double Pulsar by suitably resampling the data and obtain the time at which the modulation signal left A. This can be done by first computing the resampled time series $t_B[k]$ which represents the time of the $k^{th}$ sample measured at B, by correcting for B's orbital motion (Eq. 10 in LLW2014):
		
		\begin{equation}
		\displaystyle t_B[k] = t[k] - \frac{L}{c} - \frac{a_B \ \textrm{sin } i \ (1 - e^2) \ \textrm{sin}(\omega_B + \theta)}{c \ (1 + e \ \textrm{cos } \theta)}
		\label{tb}
		\end{equation}
		where $t[k]$ is the time corresponding to the $k^{th}$ sample measured at the solar system barycenter (ssb), $L$ is the distance to the Double Pulsar and $a_B$ is the semi-major axis of B's orbit. The $L/c$ term is a constant offset which can be neglected without loss of information. Now, we can calculate $t_A[k]$, the time when the signal causing the modulation features left A, by correcting for an additional propagation time delay along the path length from A to B:
		 \begin{equation}
		 \displaystyle t_A[k] = t_B[k] - (a_A + a_B) \ \frac{1 - e^2}{1 + e \ \textrm{cos} (\theta)}
		 \label{tA}
		 \end{equation}
		where $a_A$ is the semi-major axis of A. Eq.~\ref{tA} can be used to transform $I[k]$, the intensity data sampled at the ssb at time $t[k]$, into a frame where the time-variable Doppler shifts have been removed from the data.
		
		Using the resampled time series, we can compute a Fourier power spectrum (Eq. 19 in LLW2014):
		\begin{equation}
		\displaystyle P_{n} (z f_{A,0})_s = \left| \sum_{k} I[k] \ \textrm{exp} ( -i \ n \ \left| \Phi_{\textrm{m}} [k,z] \right|_s) \right| ^ {2}
		\label{power spectrum}
		\end{equation}
		where $z$ is a frequency scaling factor, $ P_{n} (z f_{A,0})_s$ is the power in the $n^{th}$ harmonic of the modulation corresponding to the trial spin frequency $zf_{A,0}$, and
		\begin{equation}
		\displaystyle \left| \Phi_{\textrm{m}} [k,z] \right|_s = 2\pi \ (z \ f_{A,0}) \ t_A[k] - s \ \theta(t_B[k])
		\label{phi}
		\end{equation}
		is the ``modulation phase'' which is simply the rotational (or pulsational) phase of A corrected for the sense of its rotation, with $s = 1, -1, 0$ corresponding to prograde, retrograde and no rotation (pulsation) respectively. Here, $f_{A,0}$ is the sidereal frequency of A's rotation or pulsation, which we know from timing measurements to be $44.05406$ Hz \citep{Kramer_GRtest_2006} on MJD 52997.
		
		If we compute the power spectrum in Eq.~\ref{power spectrum} for each value of $s$, then, based on the arguments in LLW2014, we should observe a peak at a frequency corresponding to $z = 1$, i.e. at $f = f_{A,0}$ and the value of $s$ with the highest power in this peak will determine the sense of rotation of A.
	
\section{Results} \label{Results}
	
	\begin{figure*}
		\includegraphics[width = \textwidth]{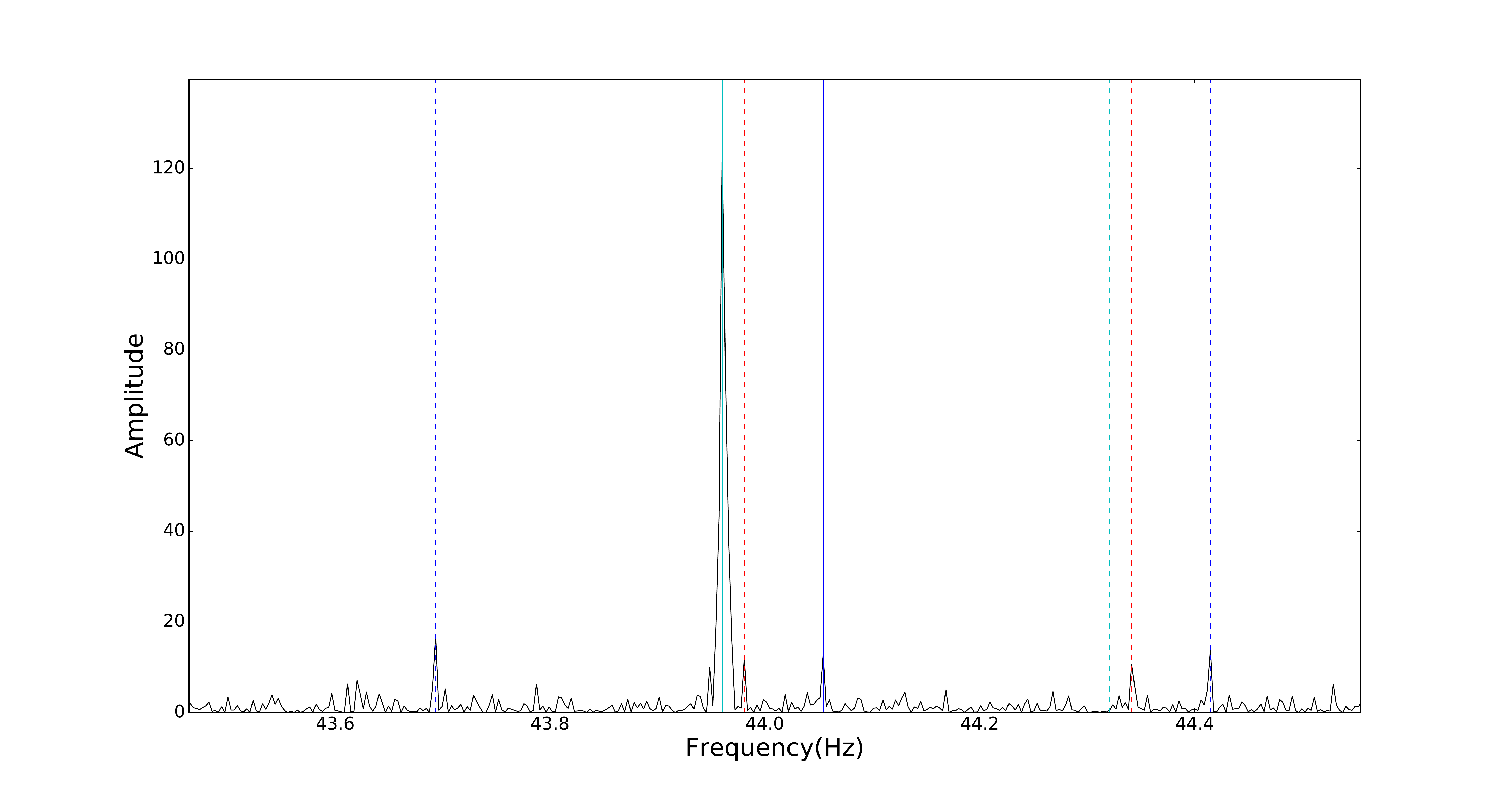}
		\caption{Fourier power spectrum for $s = 0$. Vertical dashed red lines mark the harmonics from B's intrinsic signal. The fundamental frequency of the modulation signal~($f_{A,0}$) is shown by a vertical solid blue line and vertical dashed blue lines mark the sidebands of the modulation signal. The emission from A's intrinsic signal is visible as the prominent peak close to the fundamental frequency of the modulation signal, marked by a solid vertical cyan line, which is not at $f_{A,0}$ due to the transformations applied above (see text). Similar to the sidebands of the modulation signal, we mark the positions of the sidebands for A's intrinsic signal. We can see that there is no power in the sidebands for A's intrinsic signal, indicating that we have successfully separated A's intrinsic signal from the modulation signal.}
		\label{harmonics_s0}
	\end{figure*}
	
	Since emission from A is stimulating emission in B, we can think of A as the ``carrier" signal which modulates the magnetosphere of B. This interpretation implies we would see a signal at the fundamental frequency of the carrier (in this case $f_{A,0}$) and sidebands of this signal separated by the modulation frequency (in this case $f_{B,0}$). Thus, we would expect to see a signal at frequencies $f_{A,0} \pm m \times f_{B,0}$ where $m = 0, 1, 2, ...$. We see this structure in the fourier power spectra for the three different cases of $s$, with the power spectrum for $s = 0$ shown in Fig.~\ref{harmonics_s0} for reference. We can see the peak at the fundamental frequency $f_{A,0}$ (marked by a blue solid vertical line) and its sidebands (marked by vertical dashed blue lines). We have also marked the harmonics from B's signal (fundamental frequency of $f_{B,0} = 0.3605$ Hz). They are visible as distinct peaks in the power spectrum indicating different origins for the two signals.
	
	In addition to these signals, we also see a strong signal close to $f_{A,0}$, marked in Fig.~\ref{harmonics_s0} by a solid vertical cyan line. This is the relic of the signal from A's emission, but shifted away from its native frequency of $f_{A,0}$ and reduced in amplitude by the transformations that we have applied. This is a key part of the analysis which allows us to distinguish the signal generated by A's intrinsic emission and the modulation signal. In Fig.~\ref{harmonics_s0}, we also do not detect any power in the sidebands associated with A's intrinsic signal. The presence of A's intrinsic signal, without the presence of sidebands, serves as evidence that the signal we see at $f_{A,0}$ after applying the transformations is from the modulation feature rather than from A itself, and that we have successfully managed to separate the modulation signal from A's intrinsic emission.
	
	\begin{figure*}
		\includegraphics[width=\textwidth]{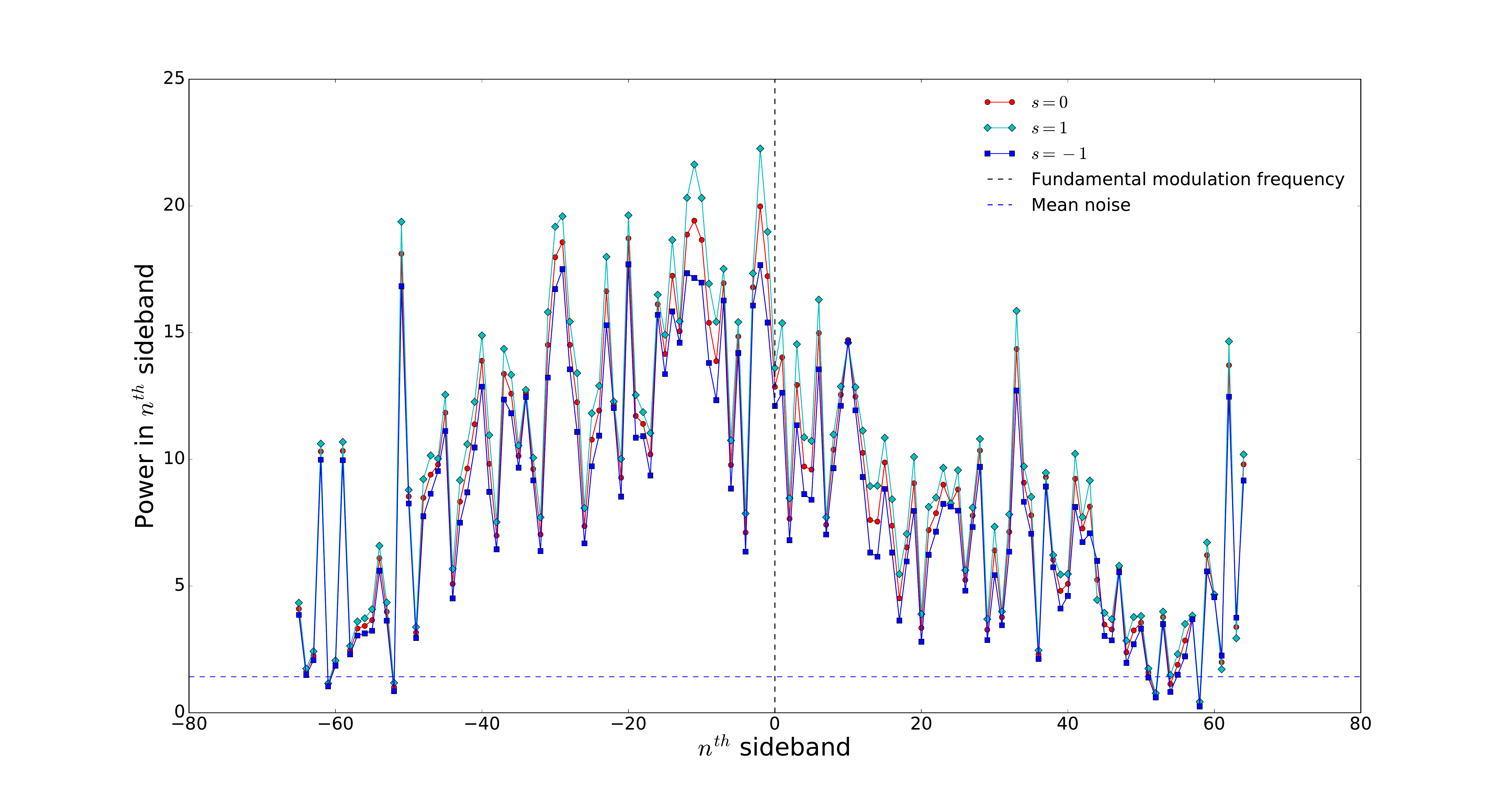}
		\caption{Fourier power in the modulation signal's fundamental frequency and sidebands for the first BP I on MJD 52997. The power for all three cases of $s = 1,0, -1$ are plotted together for comparison. The location of the fundamental frequency is shown by a vertical dashed line, while the mean noise level is shown by a horizontal dashed line. The case of $s = 1$ has consistently high power over all components which indicates this is the true sense of rotation of A.}
		\label{BP1_1_52997}
	\end{figure*}
	
	\begin{figure*}
		\includegraphics[width = \textwidth]{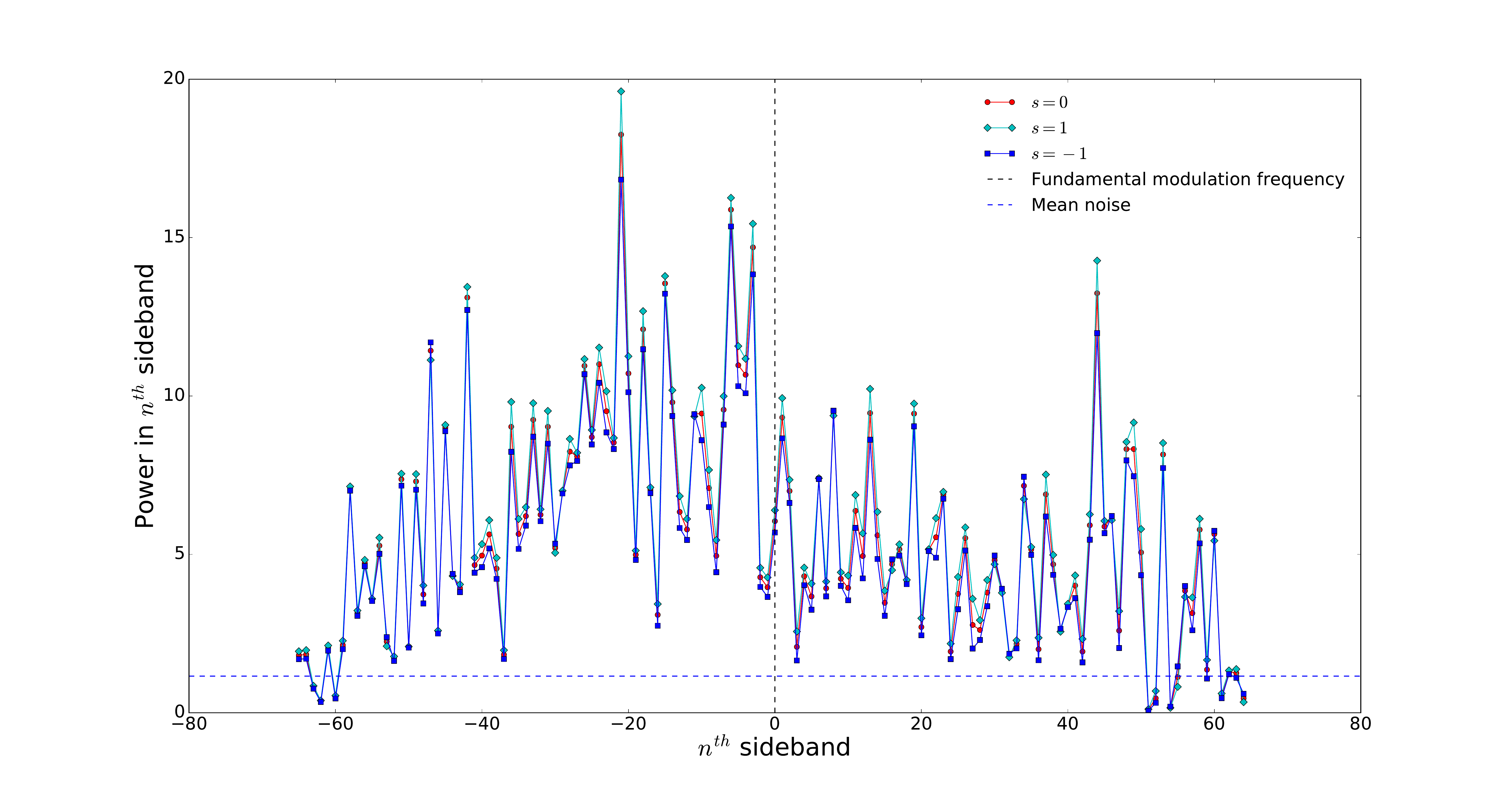}
		\caption{Fourier power in the modulation signal's fundamental frequency and sidebands for the second BP I on MJD 52997. The power for all three cases of $s = 1,0, -1$ are plotted together for comparison. The location of the fundamental frequency is shown by a vertical dashed line, while the mean noise level is shown by a horizontal dashed line. The power in $s = 1$ is consistently higher than other values of $s$.}
		\label{BP1_2_52997}
	\end{figure*}
	
	\begin{figure*}
		\includegraphics[width = \textwidth]{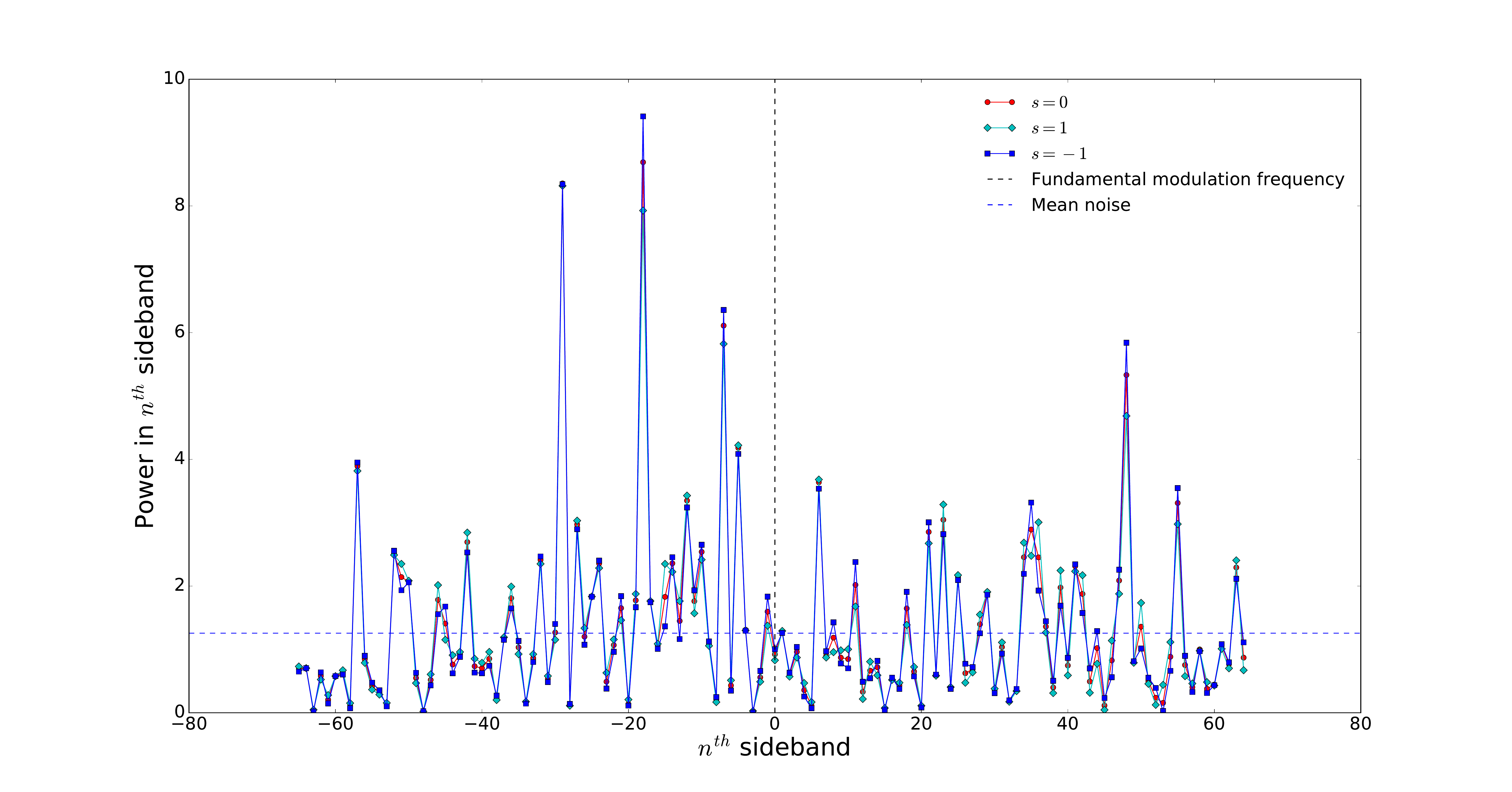}
		\caption{Fourier power in the modulation signal's fundamental frequency and sidebands for the first BP II on MJD 52997. The power for all three cases of $s = 1,0, -1$ are plotted together for comparison. The location of the fundamental frequency is shown by a vertical dashed line, while the mean noise level is shown by a horizontal dashed line. In this orbital phase range, the modulation signal is not visible. Hence, we do not see any significant power for any value of $s$ at any of the sidebands of the modulation signal.}
		\label{BP2_1_52997}
	\end{figure*}
	
	\begin{figure*}
		\includegraphics[width = \textwidth]{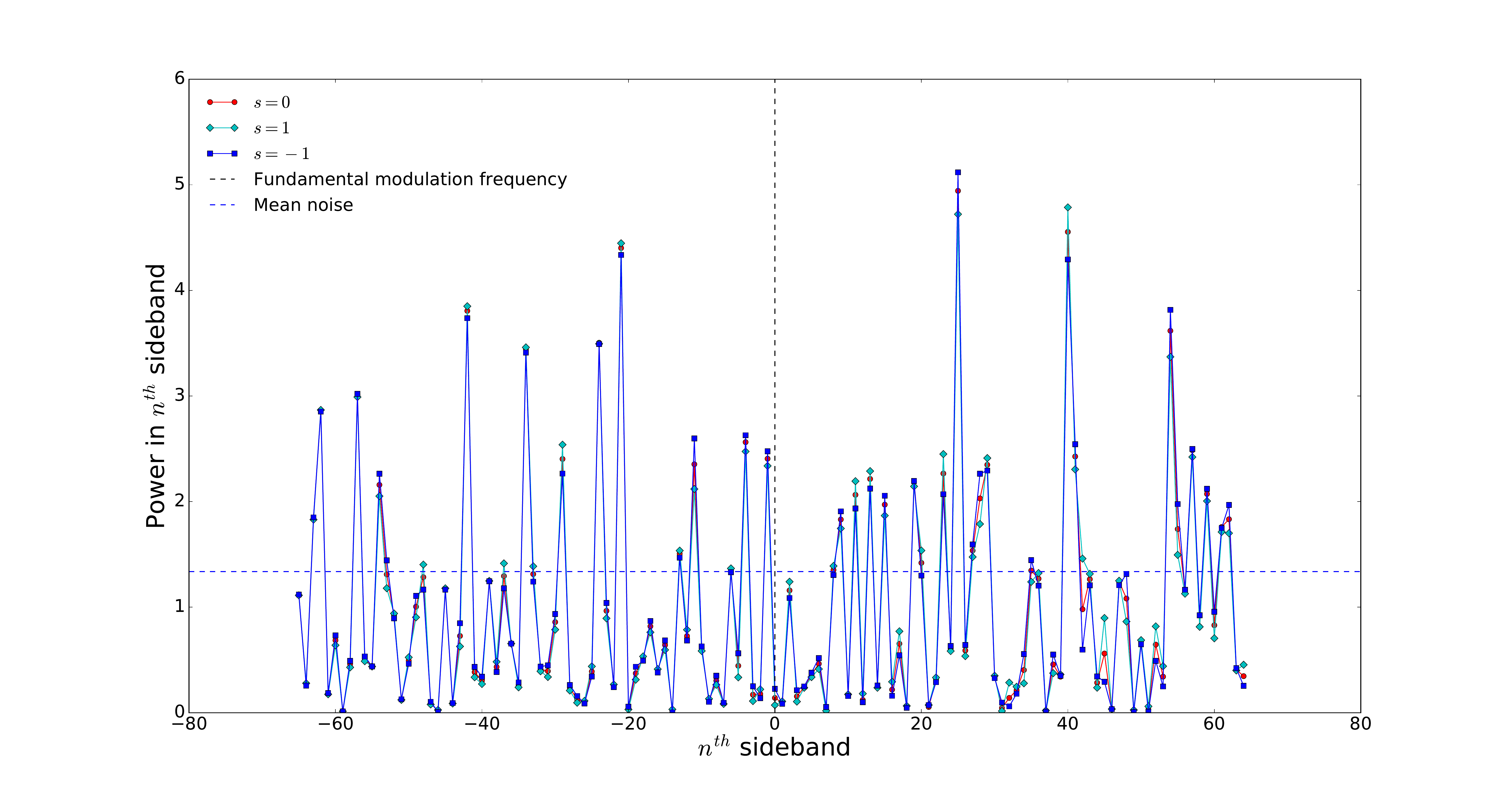}
		\caption{Fourier power in the modulation signal's fundamental frequency and sidebands for a weak phase ($40^{\circ}$ to $52^{\circ}$) on MJD 52997. The power for all three cases of $s = 1,0, -1$ are plotted together for comparison. The location of the fundamental frequency is shown by a vertical dashed line, while the mean noise level is shown by a horizontal dashed line. In this orbital phase range, B and the modulation signal is not visible. Hence, we do not see any significant power for any value of $s$ at any of the sidebands of the modulation signal.}
		\label{WP_2_52997}
	\end{figure*}
	
    Finally, we want to compare the power in the signals at frequencies $f_{A,0} \pm n \times f_{B,0}$ for all three cases of $s$, with the value of $s$ with the highest power indicating the direction of rotation of A with respect to its orbit. Note that, as described in Sec.~\ref{obs}, we have observed two complete orbits of the Double Pulsar on MJD 52997. Thus, we plot the power at the fundamental frequency and its sidebands for the BP I of the first and second orbit in Figs.~\ref{BP1_1_52997}~and~\ref{BP1_2_52997} respectively, and plot the BP II (where B is visible, but the modulation features are not visible) for the first orbit in Fig.~\ref{BP2_1_52997}. For completeness, we also plot the power at the fundamental frequency and its sidebands for some randomly selected weak phase ($40^{\circ}$ to $52^{\circ}$, where weak or no emission is observed in B's spectrum and no modulation signal is visible) for the first orbit in Fig.~\ref{WP_2_52997}.

    In both the power spectra for BP I (see Figs.~\ref{BP1_1_52997}~and~\ref{BP1_2_52997}), we can see that the fundamental frequency and its sidebands have consistently higher power in $s = 1$ than for $s = 0, -1$. The power at these frequencies is also significantly higher than the mean noise floor of the respective power spectra. By comparison, there is very little to no power in these frequencies for the other orbital phase ranges of B's orbit (see Figs.~\ref{BP2_1_52997}~and~\ref{WP_2_52997}). This is consistent with observations of the Double Pulsar where the modulation driftbands are not visible in any other orbital phase range apart from BP I. 
     
	 \begin{figure}
	     \centering
	     \includegraphics[width = \columnwidth]{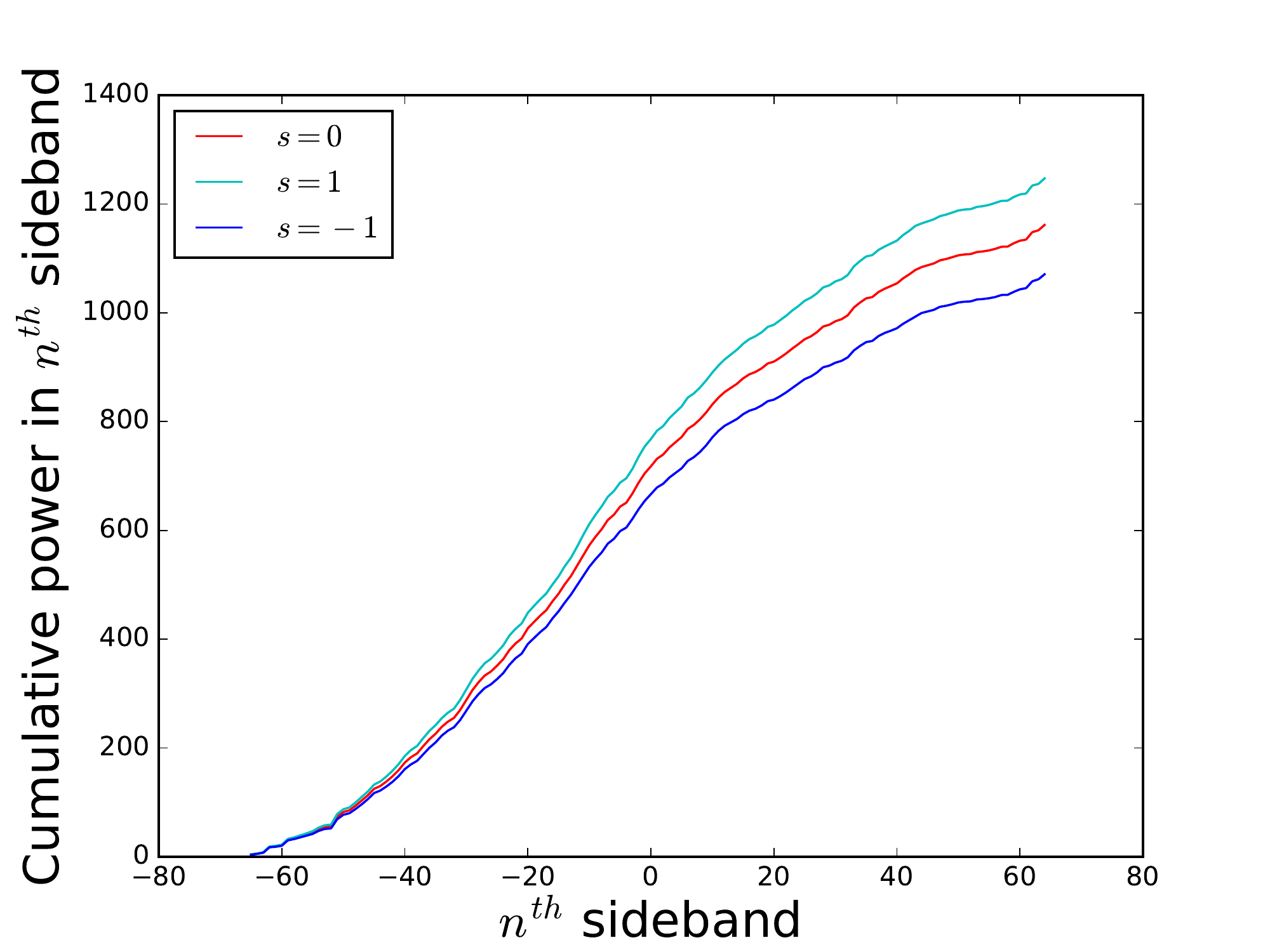}
	     \caption{Cumulative power across all sidebands for the first BP I on MJD 52997. We can see that the power in $s = 1$ is consistently higher than the power in the other values of $s$.}
	     \label{cum_bp1_1}
	 \end{figure}
	 
	 \begin{figure}
	     \centering
	     \includegraphics[width = \columnwidth]{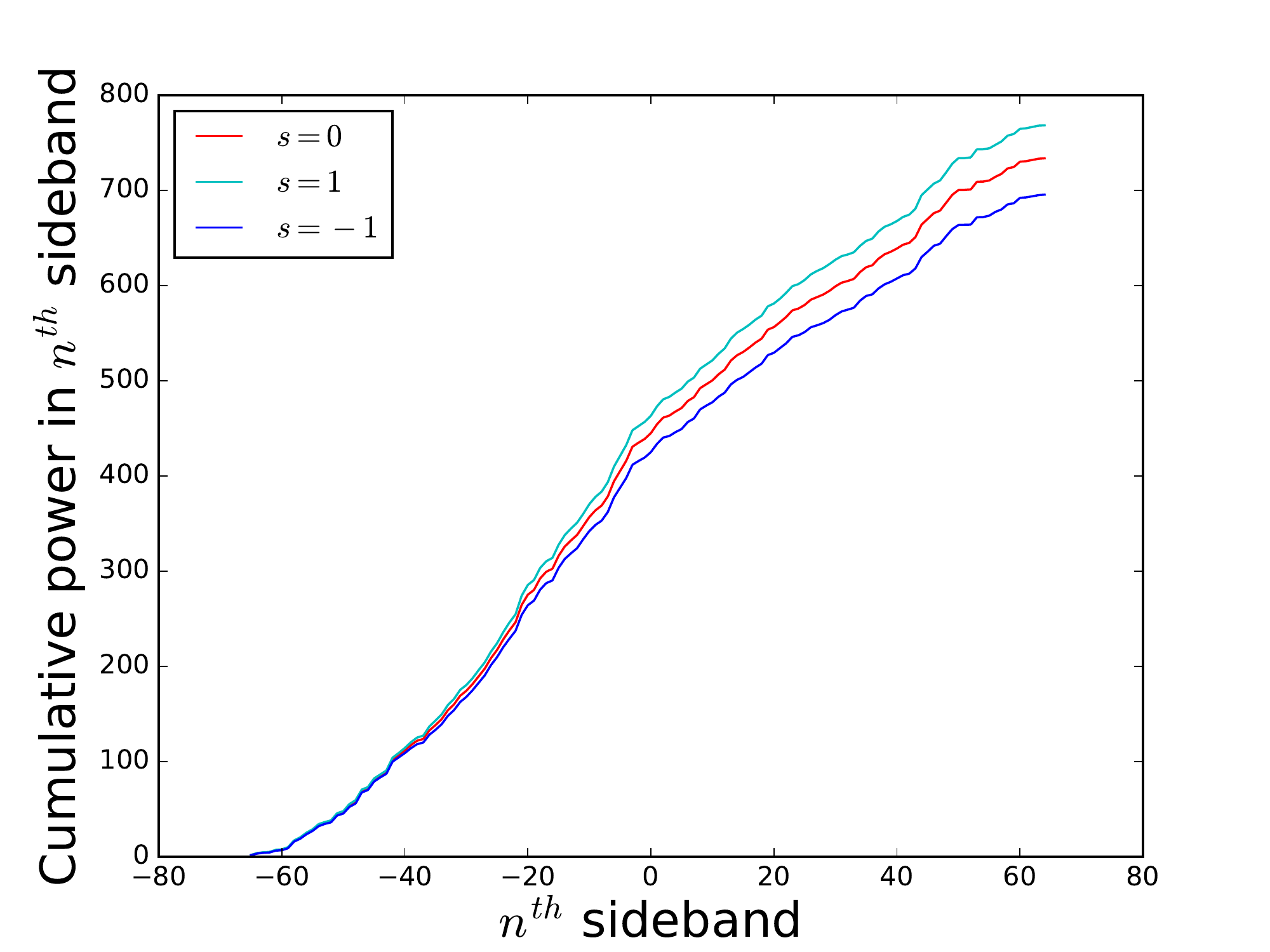}
	     \caption{Cumulative power across all sidebands for the second BP I on MJD 52997. Similar to the first BP I, we can see that the power in $s = 1$ is consistently higher than the power in the other values of $s$.}
	     \label{cum_bp1_2}
	 \end{figure}
	 
	 \begin{figure}
	     \centering
	     \includegraphics[width = \columnwidth]{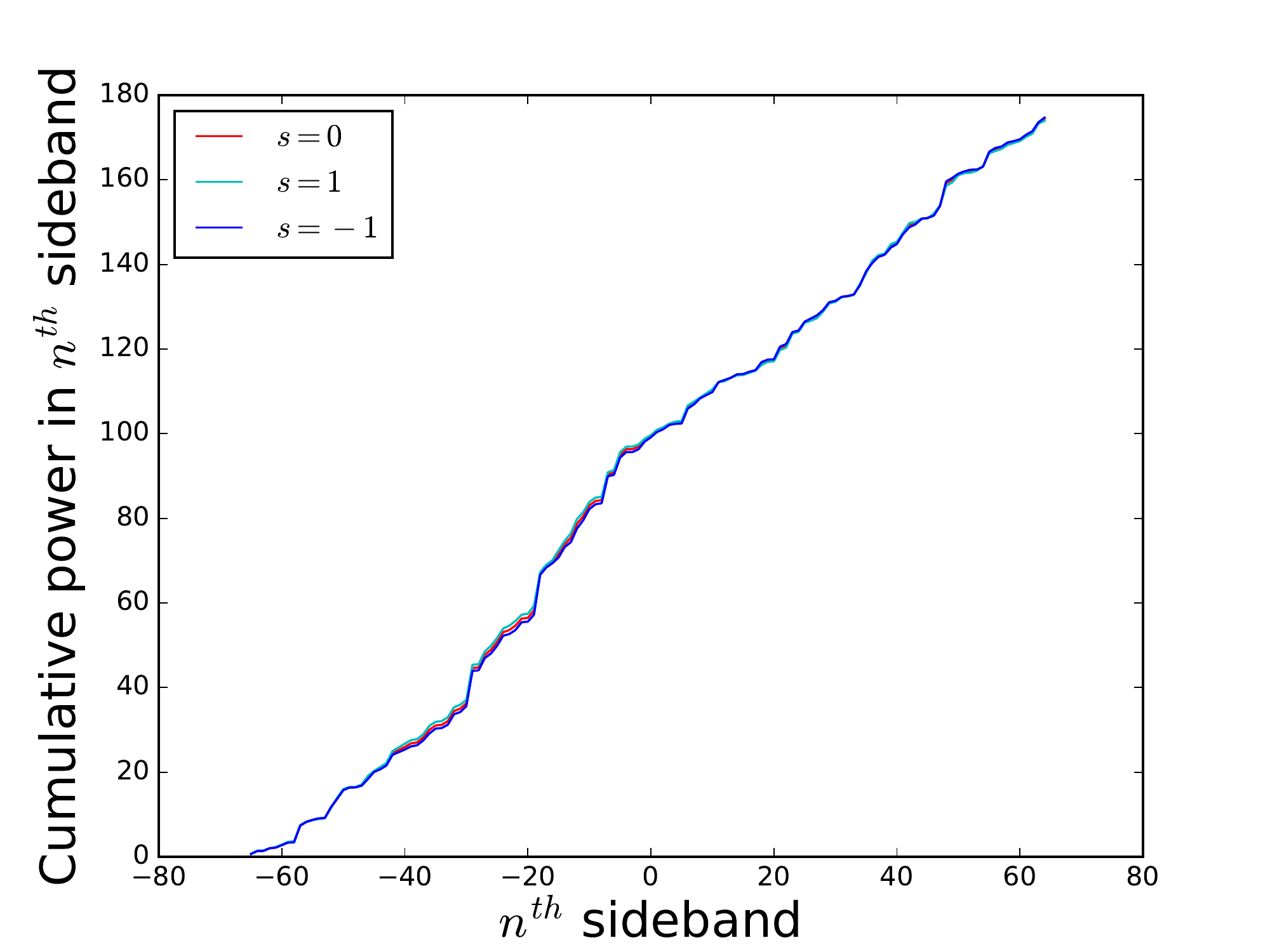}
	     \caption{Cumulative power across all sidebands for the first BP II on MJD 52997. We can see that no value of $s$ dominates over the other values in terms of total power. This is consistent with results shown in Fig.~\ref{BP2_1_52997}.}
	     \label{cum_bp2_1}
	 \end{figure}
	 
	 \begin{figure}
	     \centering
	     \includegraphics[width = \columnwidth]{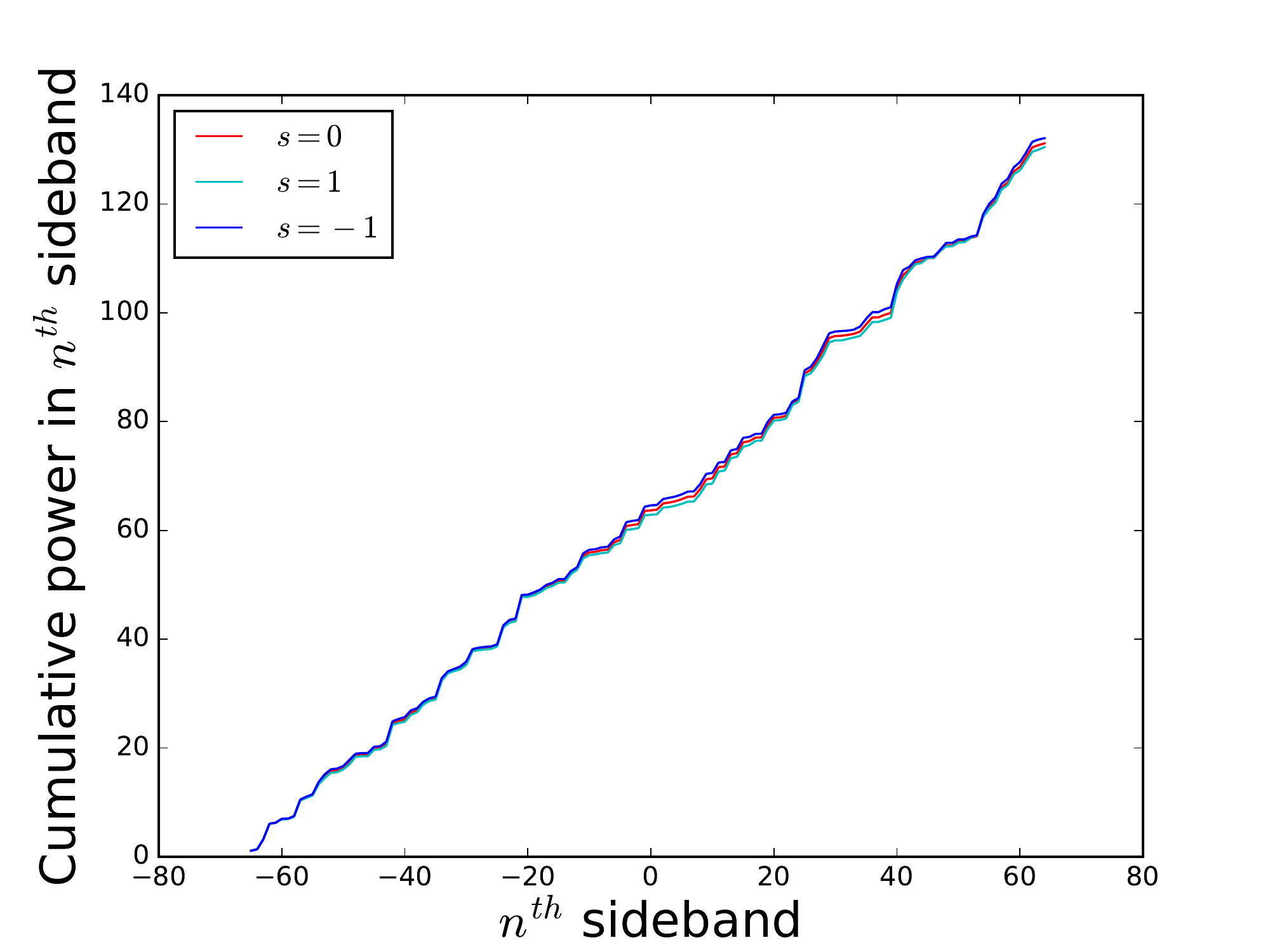}
	     \caption{Cumulative power across all sidebands for the weak phase on MJD 52997. We can see that no value of $s$ dominates over the other values in terms of total power. This is consistent with results shown in Fig.~\ref{WP_2_52997}.}
	     \label{cum_nbp_2}
	 \end{figure}
	 
	 To illustrate the consistently higher power in BP I, we plot the cumulative power across all sidebands of the modulation signal in Figs.~\ref{cum_bp1_1}~and~\ref{cum_bp1_2}. In these plots, we begin with the power in the $-64^{th}$ sideband in Fig.~\ref{BP1_1_52997} and add it with the power in the next sideband and so on until we reach the $64^{th}$ sideband. These plots clearly indicate that the power in $s = 1$ is always greater than that in other values of $s$. For comparison, we plot in Figs.~\ref{cum_bp2_1}~and~\ref{cum_nbp_2} the cumulative power across all sidebands for BP II and the weak phase respectively. Here we can see that none of the values of $s$ dominate over the other values in terms of their total power.
	 
	 To test the significance of this result, we took the data from the two BP I sections and scrambled them such that we had a time series that resembled noise. Next, we passed this scrambled time series through the same analysis described above for the real data. This process of scrambling the data and passing it through the analysis pipeline was repeated 1000 times so that we had a collection of values for the total powers in different cases of $s$. We compared these total powers across all sidebands (including the fundamental frequency) for $s = 1, 0, -1$ in the scrambled time series with the total power in $s = 1, 0, -1$ obtained from the real time series respectively to obtain the standard deviation:
	 \begin{equation}
	     \displaystyle \sigma = A \times \frac{B}{C}
	     \label{sigma_eqn}
	 \end{equation}
	 where A is the average standard deviation between total powers in the scrambled data, B is the mean of the total power for $s = 1,0,-1$ for the real data, and C is the mean of the total power for $s = 1,0,-1$ for the scrambled data.
	 
	 Using the above $\sigma$, we can compute the difference in total power for $s = 1,0,-1$ for the real data. For the first BP I, we find that $\sigma = 7.4$ and the total power in $s = 1$ is $11.6\sigma$ above $s = 0$ and $23.9\sigma$ above $s = -1$. Similarly, for the second BP I, we have $\sigma = 4.6$ and total power in $s = 1$ is $7.6 \sigma$ above $s = 0$ and $16.0 \sigma$ above $s = -1$. We perform a similar analysis for the BP II time series and find that none of the differences in total powers exceeds $1.5\sigma$. The observation of consistently higher power in $s = 1$ over $s = 0,-1$ along with the high significance of the $s = 1$ signal in two BP I time series leads us to the conclusion that $s = 1$ represents the true direction of rotation of A with respect to its orbit.

\section{Discussion and Conclusion} \label{Discussion}
	
	Based on this analysis, we can conclude that A is rotating in a prograde direction with respect to its orbit. This is the first time, in 50 years of pulsar studies, that such a direct confirmation of the sense of rotation of a pulsar has been obtained. This is additional empirical evidence for the rotating lighthouse model \citep[earlier evidence was presented by][using special relativistic aberration of the revolving pulsar beam due to orbital motion in the B1534+12 system]{stairs_1534_aberration}. This model describes pulsars as rapidly rotating neutron stars emitting magnetic-dipole radiation from their polar cap region. This rapid rotation of the pulsar resulted in the periodic pulses of light that are characteristic of pulsar emission. This work provides direct confirmation of this model.
	
	This result will help constrain evolutionary theories of binary systems \citep{Alpar_recycle_1982} as well as improve constraints on B's supernova kick. \citet[][]{Ferdman_snevidence_2013} computed a mean 95\% upper limit on the misalignment angle between the spin and orbital angular momentum axes of A to be $3.2^{\circ}$ and concluded that the A's spin angular momentum vector is closely aligned with the orbital angular momentum.This work validates that and earlier hypotheses \citep{Willems_0737formation_2006, Stairs_0737formation_2006, Ferdman_snevidence_2013, Tauris_DNS_formation} that the kick produced by B's supernova was small.
	
	Furthermore, knowing the direction of spin angular momentum of A will allow us to compute the sign of the relativistic spin-orbit coupling contribution to the post-Keplerian parameter $\dot{\omega}$, which in turn will allow us to determine A's moment of inertia \citep{Kramer_testofgravity_2009}. The moment of inertia of A, along with the well-determined mass of A will provide us with a radius, which will introduce fundamental constraints on the equation of state for dense matter \citep{Lattimer_EOS_2005}.
	
	This measurement of the sense of rotation of A was made using the frequency of the modulation signal and the rotational frequencies of A and B. An alternative way to measure the same effect is using times of arrivals of the pulses from the modulation signal and A's radio emission. \citet{Freire_model_2009} constructed a geometric model for the double pulsar system and used it to exploit the times of arrivals to measure the sense of rotation of A along with determining the height in B's magnetosphere at which the modulation signal originates. Their model will also provide another measurement of the mass ratio of A and B which will affect the precision of some of the tests of general relativity carried out in this binary system. Our preliminary results implementing the \citet{Freire_model_2009} model also indicate prograde rotation for A, and will be published in a future work. 
		
\acknowledgements

We would like to thank Joel Weisberg, Yi Liang, and Zhu-Xing Liang for useful discussions through the course of the analysis.

MAM and NP were supported by NSF award \#1517003. IHS received support from an NSERC Discovery Grant and from the Canadian Institute for Advanced Research.

\appendix

	\section{Errata} \label{errata}
		
		While performing the analysis for this paper, we discovered a typographic error in LLW2014. Eq.~10 in LLW2014 reads as:
		\begin{equation}
			t_B[k] = t[k] - \frac{L}{c} - \frac{z_B}{c}
			\label{Eq.3}
		\end{equation}
		where $t[k]$ is the barycentric time at which the $\rm k^{th}$ sample of the time series with intensity $I[k]$ was observed, $L$ is the distance to the pulsar, and $z_B$ is the projection of the position of B with respect to the binary barycentre onto the line of sight:
		\begin{equation}
		    z_B = \frac{a_B \ \textrm{sin} \ i \ (1 - e^2) \ \textrm{sin}(\omega_B + \theta)}{c \ (1 + e \ \textrm{cos} \theta)}
		    \label{zb}
		\end{equation}
		where $a_B$ is the semi-major axis of pulsar B's orbit, $e$ is the eccentricity of B's orbit, $\omega_B$ is the longitude of periastron passage for B and $\theta$ is the true anomaly.
		
		For Eq.~\ref{Eq.3} to be dimensionally consistent, $z_B$ should have dimensions of length. Looking at Eq.~\ref{zb}, we can see that $z_B$ will have dimensions of length if we remove the factor of $c$ in the denominator. Thus, Eq.~\ref{zb} should read as:
		\begin{equation}
		    z_B = \frac{a_B \ \textrm{sin} \ i \ (1 - e^2) \ \textrm{sin}(\omega_B + \theta)}{(1 + e \ \textrm{cos} \theta)}
		    \label{correct_zb}
		\end{equation}

\bibliographystyle{aasjournal}
\bibliography{bibliography}

\end{document}